# In-situ Model Downloading to Realize Versatile Edge AI in 6G Mobile Networks

*Kaibin Huang, Hai Wu, Zhiyan Liu, and Xiaojuan Qi[1]*

## Abstract

The sixth-generation (6G) mobile networks are expected to feature the ubiquitous deployment of machine learning and AI algorithms at the network edge. With rapid advancements in edge AI, the time has come to realize intelligence downloading onto edge devices (e.g., smartphones and sensors). To materialize this version, we propose a novel technology in this article, called in-situ model downloading, that aims to achieve transparent and real-time replacement of on-device AI models by downloading from an AI library in the network. Its distinctive feature is the adaptation of downloading to time-varying situations (e.g., application, location, and time), devices' heterogeneous storage-and-computing capacities, and channel states. A key component of the presented framework is a set of techniques that dynamically compress a downloaded model at the depth-level, parameter-level, or bit-level to support adaptive model downloading. We further propose a virtualized 6G network architecture customized for deploying in-situ model downloading with the key feature of a three-tier (edge, local, and central) AI library. Furthermore, experiments are conducted to quantify 6G connectivity requirements and research opportunities pertaining to the proposed technology are discussed.

## 1. Introduction

In the 1999 Hollywood blockbuster movie "Matrix", the protagonist "Neo" acquires superhuman capabilities (e.g., becoming a Judo master or dodging bullets) for his avatar in a virtual world by downloading programs from servers to his brain over wire. Though realizing such intelligence downloading to humans does not seem possible in the near future, the time for artificial intelligence (AI) downloading to edge devices (e.g., smartphones and sensors) has come. In fact, AI model downloading is one use case of edge AI being discussed for the standard of the sixth-generation (6G) mobile networks [1]. In the article, we propose a novel technology, called in-situ model downloading, that aims to achieve transparent and real-time replacement of on-device AI models by downloading from an AI library in the network. Compared with a traditional approach, its distinctive feature is the adaptation of downloading to time-varying situations (e.g., application, location, and time), devices' heterogeneous storage-and-computing capacities, and channel states.

Being AI native, 6G is expected to feature the ubiquitous deployment of machine learning and AI algorithms at the network edge, which are collectively known as edge AI [2]. AI models with relatively small sizes can be completely downloaded onto devices to enjoy the advantages of better data security, faster decision time, context and location awareness, and a lighter burden on uplinks. This is supported by the latest generation of mobile chips designed by leading semiconductor companies, for example, Qualcomm, Apple, Samsung, Huawei, and NVIDIA. They share the common feature of comprising powerful graphics processing units (GPUs) or other AI acceleration hardware to support the training and execution of AI models. The hardware will endow edge devices capabilities in natural language processing, image recognition, and video content analysis, which lays a platform for implementing intelligent IoT applications. Nevertheless, the sole reliance on mobile hardware to implement on-device AI is confronted by the conflicts between the hardware's limited storage and computation resources and the large sizes of high-performance models. Furthermore, numerous application-specific models are needed for a device to support a matching number of applications or adapt to changes in time, locations, and context, or users' preferences and behaviors.

[1] The authors are with Dept. of Electrical and Electronic Engineering, The University of Hong Kong, Hong Kong. Contact: K. Huang (email: huangkb@eee.hku.hk).

These issues can be addressed by on-demand model downloading from an AI library in the cloud to meet a device's real-time needs. A practically unlimited number of models can be kept in the AI library and managed by grouping them into different categories according to service types, environment, user preferences and requirements, and hardware specifications. Thereby, all possible needs of AI by devices can potentially be met. The deployment of on-demand model downloading in 6G networks is still at a nascent stage and faces two main challenges among others.

- **In-time downloading.** To avoid interrupting ongoing applications, downloading has to meet an application-specific latency requirement. 6G aims to use AI to empower wide-ranging applications of tactile communications — augmented/virtual reality (AR/VR), remote robotic controls, and auto-navigation, to name a few [3]. Such applications demand end-to-end latency to be as low as several to tens of milliseconds.

- **Devices' heterogeneous capacities.** One aspect of heterogeneity in edge devices is their distribution over a broad spectrum of computing-and-storage capacity with tablets and smartphones at the high-end and passive RFID/NFC tags at the low-end. Another aspect of heterogeneity is reflected in devices' communication capacities determined by available radio resources (e.g., array size, transmission power, bandwidth, and coding/decoding complexity), channel states [4], and potential interference from neighboring devices [5]. The above heterogeneity requires adaptation of the size of a model being downloaded to the targeted device' hardware and channel.

Other challenges include, for example, the need of local model fine-tuning, device-server cooperation, and predictive downloading. The use of existing 5G technologies to support model downloading is inefficient as they are not task-oriented and hence lack the desired high level of versatility and efficiency, the native support of heterogeneity, and a guarantee on end-to-end performance.

To answer the call for developing an advanced technology to tackle the mentioned challenges, we propose the framework of in-situ model downloading. To adapt to devices' heterogeneous hardware constraints and link rates, we first propose three approaches, termed depth-level, parameter-level, and bit-level in-situ model downloading, by building on existing techniques from early exiting for inference, model pruning, and quantization (see Section 2). The approaches enable adjustments of the number of layers and parameters and the level of precision of a model to accommodate devices' heterogeneous requirements. It is possible to integrate these three approaches to generate a large-scale AI library, which comprises high-granularity models, to support a versatile downloading service. Second, we propose a 6G network architecture to implement in-situ model downloading with key features, including a three-tier (edge, local, and central) AI library, cooperative network management by operators and service providers, task-oriented communications, and mobile architecture for transparent downloading (see Section 3). Third, we conduct experiments to quantify 6G connectivity requirements for realizing in-situ model downloading (see Section 4). Last, we conclude with a discussion on other research challenges for the new technology and potential solutions (see Section 5).

## 2. Techniques for In-situ Model Downloading

As mentioned earlier, the large population of edge devices exhibits a high level of heterogeneity in different dimensions. This requires flexible methods for an in-situ model generation to accommodate the heterogeneity and channel adaptive downloading to cope with time variations of link rates. In the following sub-sections, we build on diversified techniques from the areas of split inference, early exiting, and model compression to propose three techniques, namely depth-level, parameter-level, and bit-level in-situ downloading. Though in different ways, they all enable a mobile model to be generated in real-time (or retrieved from a pre-generated AI library) and downloaded onto a device based on its feedback of *device situation information* (DSI), which specifies its capacities, hardware types, channel states, location, and accuracy-and-latency requirements. It is even possible to integrate the three techniques to achieve high granularity in model generation and downloading as illustrated in Fig. 1. Through adopting the three-level compression, the processed in-situ model downloading promises to meet the real-time requirements of edge intelligence applications on inference speed, energy consumption, prediction accuracy, etc.

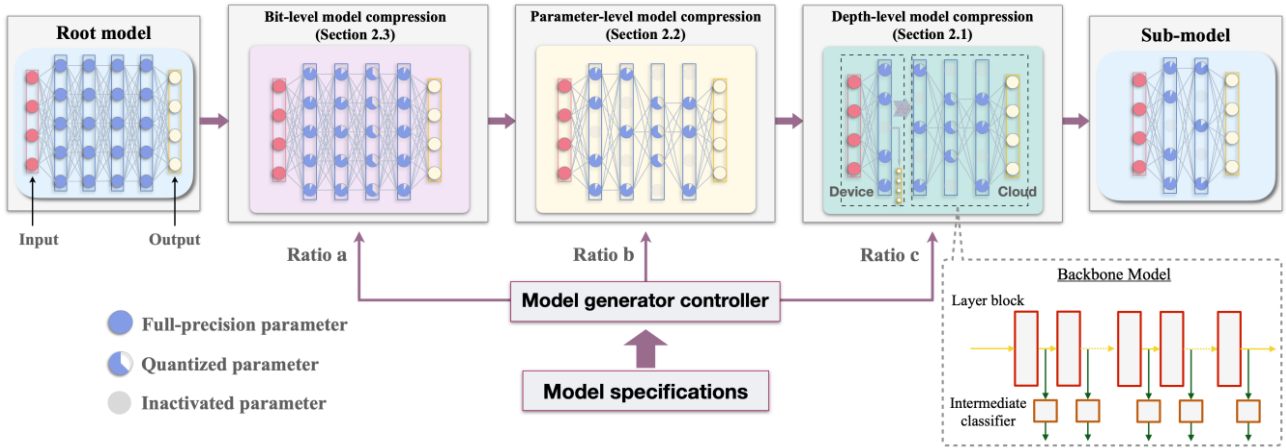

Fig. 1. The integration of the techniques of bit-level, parameter-level, depth-level in-situ model downloading to enable high-granularity real-time sub-model generation or AI library construction.

## 2.1 Depth-Level In-situ Model Downloading

Split inference refers to the class of techniques that divides a root deep neural network model into the first and the second halves for deployment at a device and a server, respectively [3]. This requires uploading of features output by the former, called a local model, for feeding into the server model for generating the inference result. This leverages server computation resources to alleviate the device's computation load while preserving its data privacy. The splitting point can be adjusted for the purpose of load balancing [6]. In the context of in-situ model downloading as illustrated in Fig. 1, we propose to adapt the splitting point, or equivalently the depth of the mobile model being downloaded, to the device's DSI feedback for real-time model generation. Alternatively, sweeping the splitting point across a root model generates a set of mobile models with complementary server models in the AI library. The advantages of the depth-level method are threefold. First, its support of device-server cooperation implies that even when the device's hardware or radio resources are limited such that the downloaded model is small, a high inference accuracy can be achieved with server assistance. Second, the depth-level model downloading makes it easy to implement channel adaptive transmission, where the number of downloaded layers is adapted to the downlink rate, or layer-by-layer progressive transmission. Third, the uploading of features offers benefits for protecting data privacy at a level that increases as the number of downloaded layers grows.

There exists a downlink-uplink tradeoff for the proposed depth-level technique. This results from the well-known fact that for many popular models (e.g., auto-encoder, VGGNet, MobileNet, and the latest

ConvNeXt), the size of intermediate features (in bits) shrinks with an increasing depth that a task traverses a deep neural network [7]. In the current context, the tradeoff is translated into that the growth in the downloaded model size leads to the reduction of uplink data size (or equivalently communication latency) for server inference. Consequently, it is desirable to increase the size of a downloaded model to rein in the uplink overhead when there are many data samples for inference, or otherwise decrease the size to reduce the downlink overhead.

We propose an enhanced version of depth-level in-situ model downloading using a model architecture called *backbone neural network*. As illustrated in Fig. 1, the model comprises a conventional deep neural network added with multiple low-complexity intermediate classifiers to allow early exiting of a task from the model upon reaching a target inference accuracy (or confidence level). As a task traverses deeper into the model, the feature is abstracted to a higher level through the convolution and pooling operations, making different classes easier to differentiate and thus increasing the classification accuracy [7]. However, executing more layers induces high computational latency. This induces an accuracy-latency tradeoff useful for supporting tasks with heterogeneity in performance and latency requirements. The use of such a model in the depth-level technique yields the flexibility of early exiting either on a device or a server depending on the task requirements.

## 2.2 Parameter-Level In-situ Model Downloading

An alternative technique, termed parameter-level in-situ model downloading, is to adapt the number of model parameters based on *network pruning,* referring to reducing the size of a neural network model by pruning its *least important* parameters [8]. Conventionally, network pruning is performed for reducing the network complexity, avoiding model over-fitting, or discovering a good sub-model architecture embedded in the root model. In the current context, a model is pruned for a new reason - to adapt to the DSI of a device requesting downloading. The specific procedure consists of the following steps. First, a device sends a downloading request together with DSI to the network. Second, this triggers an associated edge server to determine the target dropout rate based on DSI and uses it to prune a root model accordingly using a chosen importance metric (to be discussed in the sequel). Third, the pruned model is re-trained to prevent the model from suffering dramatic performance loss from pruning. Last, a prepared model satisfying accuracy and latency requirements is downloaded onto the device. Typically, the latency and energy consumption of model pruning and re-training is negligible due to the sufficient computation resources at the edge server. In latency-critical scenarios, variations of the original model with different parameter-pruning ratios can be trained in advance and stored in the server's memory such that a suitable model can be retrieved and downloaded instantaneously in response to a request. In parameter-level model downloading, model downloading can be made even more versatile by enabling progressive downloading. This allows a downloaded model to be integrated with a previously downloaded one to generate a larger model outperforming its two individual components. Implementing such a technology requires a sophisticated subnet generation technique, called "all-in-one", that simultaneously optimizes multiple subsets [9]. As a result, each subnet is embedded in any other one larger in size. Then differential parameter sets can be defined to facilitate incremental downloading over multiple channel uses.

Compared with the preceding layer-based technique, the current one as well as the bit-level technique in the sequel provide finer granularity over the ranges of accuracy and model sizes but is not streamlined for device-server cooperation.

## 2.3 Bit-Level In-situ Model Downloading

As mentioned, edge devices (e.g., IoT sensors) are usually under many hardware constraints in terms of, for example, low-precision logic circuits, and limited memory and energy. Therefore, as discussed in the sequel, there exists rich literature on techniques for quantizing neural network models (or their parameters to be more precise) so as to increase inference speed and reduce energy consumption at edge devices. The proposed bit-level in-situ model downloading is to adapt the parametric bit-width and size (total bits) of a downloaded model according to the requesting device's hardware requirements. For this approach, upon receiving a device's request with DSI, a quantized model with a matching resolution and size can be either generated from a high-precision (i.e., floating point) one in real-time or retrieved from a pre-generated AI library in the network. The bit-level in-situ model generation can leverage a wide range of existing techniques and methods [10]. Perhaps the simplest procedure is directly quantizing a model (using a uniform/non-uniform, scalar/vector quantizer) followed by fine-tuning the quantized parameters by re-training. Note that similarly as in pruning, re-training is important to reduce performance loss due to quantization. Researchers have found that allowing model parameters to have different precisions provides a way to avoid significant performance degradation at a low-precision level, leading to the development of mixed-precision techniques. Typically, different layers in a model are quantized with varying bit precisions, which are optimized to match the layers' importance levels in terms of their effects on inference accuracy [11].

It is worth mentioning that alternative model compression techniques such as neural architecture search (NAS), model partitioning, and multi-exit DNN can be adopted to streamline model compression at different levels [9, 12]. For instance, the existing NAS algorithms can be applied to the search for the best layers/parameters/bits to keep for depth/parameter/bit-level compression, respectively. For the compression techniques adopted, on-device fine-tuning can further boost the AI performance by customising the model parameters for heterogeneous settings.

## 3. Network and Mobile Architectures

Several new features of 6G networks facilitate the realization of in-situ model downloading [2]. A distinctive feature of 6G is the spreading of AI services from the central cloud to every corner of the network. Bringing AI to the proximity of users not only achieves superior performance (e.g., ultra-low latency) but also helps to protect users' data privacy. On the contrary, 5G does not consider edge AI as a native feature, let alone providing the infrastructure to deploy models targeting heterogeneous latency requirements from the central cloud to the network edge. The second feature of 6G is the paradigm shift of air-interface technologies from being *rate-centric* as in 5G networks, namely aiming to support high-rate "bit pipes", to being *task-oriented*, which focuses on the efficiency and effectiveness of executing a specific task. Implementing model downloading on the former paradigm of 5G is, although feasible, far from optimality in terms of end-to-end task performance metrics, e.g., the accuracy performance of the downloaded model, since the downloading process decoupled from the radio access layer. On the contrary, task-oriented communications in 6G for in-situ model downloading will advocate communication-computing integrated design to provide the desired versatility and maximum efficiency in accommodating heterogeneity in devices' constraints and users' requirements. Another feature of 6G pertaining to our discussion is that its virtualized infrastructure pools the resources of and allows joint management by operators, users, and (third-party) service providers. In particular, the operators and service providers will jointly manage the AI libraries stored in the network and their uses to support the model-downloading service. A model-downloading service can be implemented using end-to-end network slicing. The resultant virtualized network is illustrated in Fig. 2; its main components and operations are described as follows.

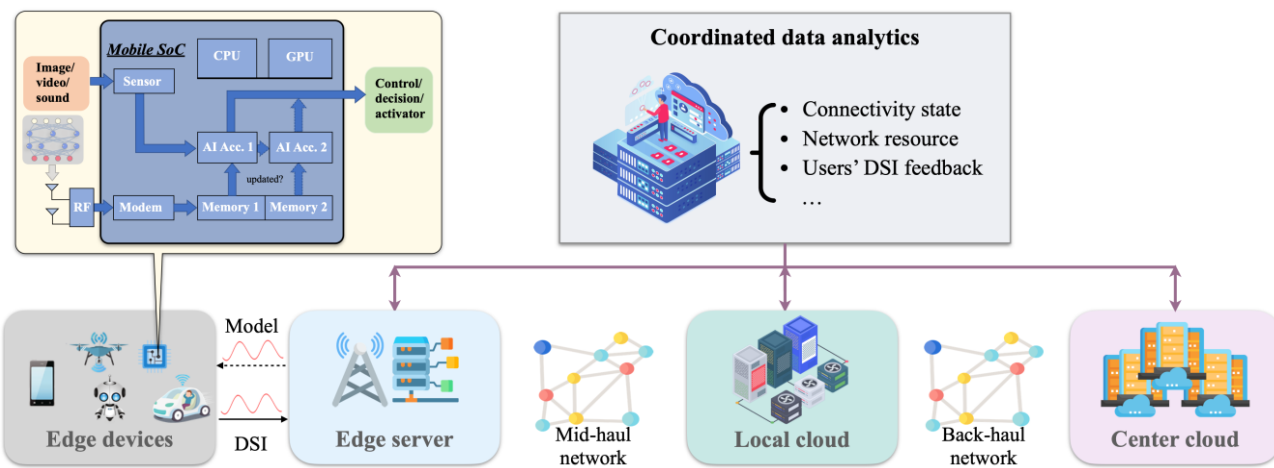

Fig. 2. Network and mobile architectures for deploying the service of in-situ model downloading.

First, an AI library that is built and managed by a service provider for a specific service can be stored in the *multi-layer cloud* of a 6G network, which comprises an edge cloud (formed by edge servers and base stations), local cloud, and central cloud as illustrated in Fig. 2. The models in the library can be separated into three types, namely real-time models, non-real-time models, and generic models. They are stored in the edge, local, and central clouds, respectively. Real-time models are task/application-specific models for latency-sensitive services such as AR/VR, autonomous driving, online gaming, and robotic controls. Their placement in the proximity of users is essential for ultra-low latency downloading to support tactile applications. On the other hand, non-real-time models are task/application-specific models for services having minimum real-time requirements, for example, task-specific object recognition for images and videos. Downloading such models from a local cloud involves multi-hop transmission across a mid-haul network (see Fig. 2). Unlike the two preceding model types, generic/foundation models trained on standard datasets can be downloaded offline for applications without real-time requirements; for example, downloading a VR model for animal classification to a headset before its use in a biology class. In particular, a foundation model is a large-scale generic model (e.g., GPT-3 with 175 billion parameters) that can be adapted to a wide range of downstream tasks. While downloading latency is not the most critical issue for non-real-time and generic models stored at local and center clouds, the model size still needs to adapt to communication budget, local storage capacity and device accuracy requirements as specified by the DSI. The enormous population/sizes of generic/foundation models can be accommodated by practically unlimited space in the central cloud. In building the AI library, each mentioned type of model can be further categorized in other dimensions including model architecture, size, and parametric resolution based on model generation techniques discussed in Section 2.

Second, the cooperative network management by operators and service providers and task-oriented communications in 6G networks will facilitate the communication-and-computing integrated design of in-situ model downloading. In particular, the 6G networks will provide mechanisms for a service provider to monitor connectivity state (i.e., data rate, latency, and reliability) and network resource utilization as well as receive users' DSI feedback. For implementation on the network layering architecture, this requires message passing between the application layer on the top and the radio-access layers at the bottom. The access to real-time system/user state information enables the service provider to prepare a suitable model in response to a downloading request, which is retrieved from an in-cloud AI library or generated in real time. Furthermore, tracking a user's location helps the service provider to predict the time of model use and the corresponding connectivity state so as to perform downloading in advance and make the operation intelligent and transparent. For instance, the network can predict the time a user enters a car or a museum and download an auto-navigation model in the former case and switch it in time to a tour-guide model to ensure uninterrupted user experiences. In addition, a task-oriented design for in-situ model downloading can involve the joint design of adaptive transmission techniques (e.g., adaptive modulation and coding and MIMO beamforming) and adaptive model downloading for an end-to-end performance metric (e.g., inference latency and accuracy).

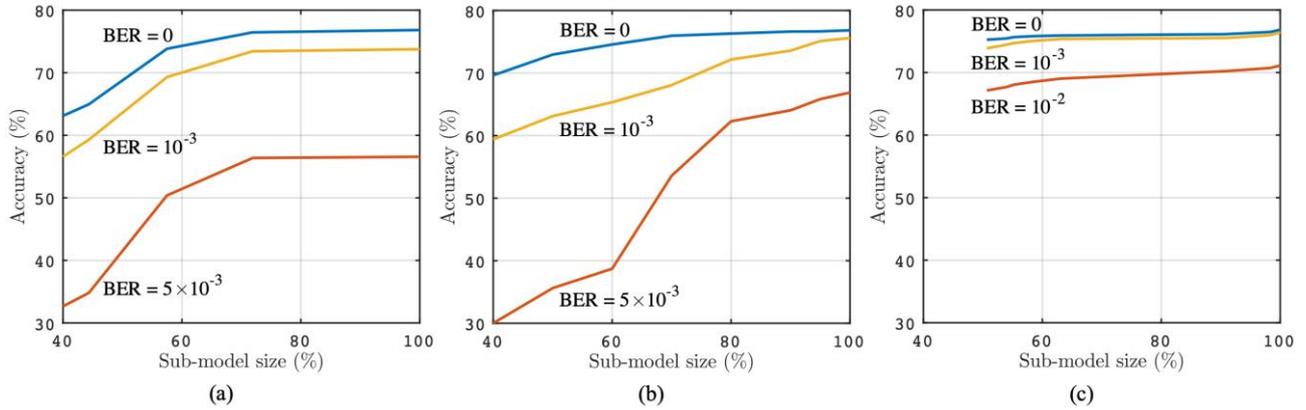

Fig. 3. Local inference accuracy versus downloaded model size under reliable (BER = 0) and unreliable (BER > 0) links with model compression in: (a) depth-level; (b) parameter-level; (c) bit-level. For bit-level model compression, the sub-model size is evaluated against that of a well-trained 4-bit quantized model.

Third, the network function of *coordinated data analytics* as illustrated in Fig. 2 is to perform data analytics of user experiences by leveraging continuously generated user feedback and data aggregated across networks. The results can help service providers and operators to improve the AI models using more training data and techniques such as reinforcement learning and streamlining network operations for the model-downloading service.

Last, the proposed mobile architecture for in-situ model downloading is illustrated in Fig. 2. Aligned with the current trend, the mobile architecture supporting in-situ model downloading comprises different types of cores, namely central processing units (CPU), graphics processing units (GPU), and AI accelerators, to support diversified types of computing tasks and applications. While CPUs are used for generic applications and controlling GPUs and AI accelerators, GPUs support graphic processing and gaming. On the other hand, AI accelerators, which are of our interest, are application-specific integrated circuits dedicated to accelerating AI and machine learning without interrupting operations in other parts of the system-on-chip. To ensure transparent model downloading, the proposed architecture features two AI accelerators with two associated memory units. This allows downloading of a new model from the wireless receiver into Memory 1 and then loading the model into AI Accelerator 1 while the current model is executed in parallel using Accelerator 2 and Memory 2. The models are switched upon completion of the new model downloading and loading.

## 4. 6G Connectivity — Experiments and Requirements

In this section, experiments are conducted to investigate the requirements of 6G connectivity to deploy in-situ model downloading. First, we study the reliability requirements. The three techniques proposed in Section 2 are compared for the task of image classification using the popular mobile-model architecture, ResNet-18, with 11.69 million parameters and trained on the widely used dataset of CIFAR-100. The parameters of the ResNet-18 root model are in the floating-point format (32-bit precision). For depth-level in-situ model downloading, an intermediate classifier is appended to the downloaded block of network layers to allow immediate on-device classification without server assistance. For parameter-level compression, we prune the well-trained network parameters according to their magnitude. On the other hand, the bit-level technique is implemented by quantizing each model parameter into a 4-bit representation followed by mixed-precision bit allocation over layers [11]. The downloaded model data consist of *topology information*, which defines the model architecture, and *model parameters*. The former is much smaller in size (e.g., 1 Mb) than the latter (e.g., 800 Mb for ResNet-50) but more critical. Therefore, topology information is assumed to be transmitted over a reliable channel while that for transmitting model parameters is prone to bit errors with reliability specified by a bit-error rate (BER). In Fig. 3, we present the accuracy performance of local inference against downloaded sub-model size (in percentage of root-model size). Both the cases of reliable (BER = 0) and unreliable links are considered. One can observe from Fig. 3 that all in-situ model downloading techniques in comparison are robust against model compression that confirms their practicality. In terms of required reliability, it is found that a BER of $10^{-3}$ is sufficient to rein in the performance loss with respect to the ideal case. This suggests that for in-situ model downloading, ultra-reliable communication (with a BER usually below $10^{-9}$) is an overkill and normal reliability is sufficient due to AI models' robustness against parameter loss or inaccuracy.

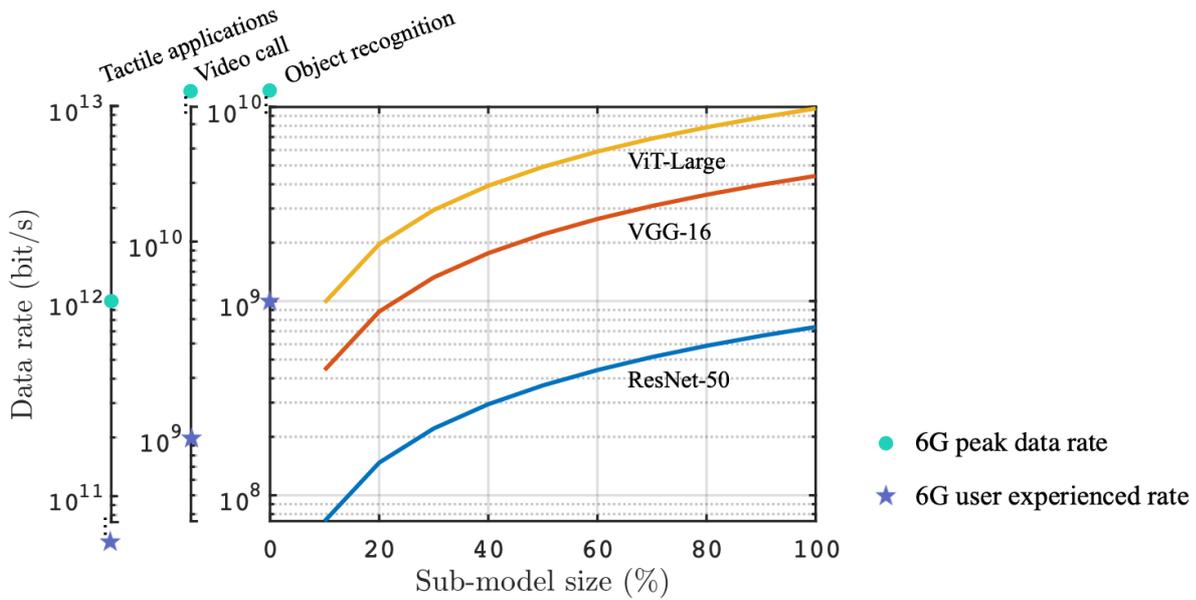

Fig. 4. Data rates required for downloading different types of AI models under application-specific over-the-air latency.

Next, we investigate the data-rate requirement for in-situ model downloading under an application specific latency requirement. We consider three applications with increasingly latency sensitivity, namely image recognition, video call, and tactile communications, with allowed over-the-air downloading latency of 1s, 200 ms, and 1ms, respectively [1]. We consider three popular mobile AI models ranging from small to large in size. They are ResNet-50, VGG-16, and ViT-Large (Vision Transformer), with 23 million, 135 million, and 307 million parameters, respectively. They are considered as root models from which a sub-model is generated (using a technique in Section 2) for downloading. The curves of data rates required for a varying size of sub-model are plotted in Fig. 4.

Consider an image-recognition task. From the figure, we can observe that ResNet-50, which is relatively small in size, can be deployed in-situ with a link providing the 6G user experienced rate for any sub-model size. Here, the user experienced rate refers to the minimum data rate required to achieve a sufficient quality experience, projected to reach 1 Gb/s in 6G networks [13]. As for the larger VGG-16 model, meeting the latency requirement need removing 80% of the total bits to cut down the required data rate to 0.86 Gb/s. However, for the state-of-the-art ViT model, the user experienced rate is insufficient as the required rate is up to 10 Gb/s for full-model downloading. Even downloading 20% of the model requests a data rate of 1.96 Gb/s, which exceeds the 6G user experienced rate. Next, consider the application of video call. The user experienced rate is too low to support in-situ downloading except for ResNet compressed to about 30%. Last, for applications of tactile communication, the peak rate is sufficient only for downloading a small model like ResNet. Additionally, one can observe that the 5G user experienced rate (i.e., 100 Mbps) fails to support the model downloading in almost all cases.

## 5. Concluding Remarks

The proposed framework of in-situ model downloading supports edge devices to adapt their AI models to changes in the application, environment, and user requirements so as to maintain high effectiveness of applications and satisfactory performance in inference and decision making. At the same time, the framework features versatility by accommodating devices' heterogeneous hardware and connectivity constraints by supporting flexible partial model downloading and device-server cooperation. In-situ model downloading will support wide-ranging AI-empowered applications in 6G including face and speech recognition, multimedia editing, VR/AR, machine-human interface, auto-driving, recommendation, and personalization in e-commerce, advertisement, and finance. For instance, downloading multimedia-editing models can enable devices to perform automatic object tagging, scene interpretation, and audio/visual quality enhancement; surveillance models can enable cameras to detect traffic accidents and offences, predict emergency situations, and control traffic.

Being a largely uncharted area, in-situ model downloading provides a goldmine of research opportunities. Several are described as follows.

- **Device-server cooperation.** Such cooperation is not native to parameter-level or bit-level model generation techniques in Section 2. Cooperation techniques can be developed for improving the communication efficiency, which allows locally extracted features to be reused for server inference when server assistance is needed.

- **Design with graceful degradation.** Models being downloaded can encounter missing parameters or layers due to packet loss caused by link unreliability. The issue can be addressed by designing robust models maintaining essential or limited functionality even in the case of losing large portions by replacing them using local versions.

- **Model customisation.** The proposed model downloading techniques can be enhanced with feature of model customisation, for which a downloaded model is fine-tuned using local data for further training on a downstream task. Using a different method, called muNet, a downloaded deep neural network can be customized by inserting new layers or removing selected layers using an evolution algorithm [14].

- **Distributed model improvement.** In return for downloading and using a shared AI model, users can contribute to improving the model via server coordinated federated learning using local data or distributed reinforcement learning via users' interaction with the physical world [15]. This gives rise to opportunities for integrated design of model downloading and distributed learning.

- **Predictive model downloading.** Useful algorithms can be designed for on-device or on-server deployment to detect the mismatch of a current model with the changing context (as reflected in, e.g., degrading inference performance or confidence level) and trigger downloading of a new model or model update.

## 6. Acknowledgement

The work of K. Huang described in this paper was substantially supported by a fellowship award from the Research Grants Council of the Hong Kong Special Administrative Region, China (Project No. HKU RFS2122-7S04). The work was also supported by Guang-dong Basic and Applied Basic Research Foundation under Grant 2019B1515130003, Hong Kong Research Grants Council under Grants 17208319, and Shenzhen Science and Technology Program under Grant JCYJ20200109141414409.

# Biographies

**Kaibin Huang** [F] (huangkb@eee.hku.hk) is a Professor at the Dept. of Electrical and Electronic Engineering, The University of Hong Kong, Hong Kong. He has been named as a Highly Cited Researcher by the Clarivate Analytics in 2019-2022. He is an Executive Editor of IEEE Transactions on Wireless Communications, and an Area Editor for both IEEE Transactions on Machine Learning in Communications and Networking and IEEE Transactions on Green Communications and Networking. He was an IEEE Distinguished Lecturer.

**Hai Wu** (wuhai@eee.hku.hk) received the B.Eng. degree from Southern University of Science and Technology, Shenzhen, in 2020. He is currently working towards the Ph.D. degree with Dept. of Electrical and Electronic Engineering, The University of Hong Kong (HKU), Hong Kong. His recent research interests include deep learning and efficient deployment of artificial intelligence.

**Zhiyan Liu** (zyliu@eee.hku.hk) received the B.Eng. degree from the Dept. of Electronic Engineering, Tsinghua University, Beijing, in 2021. He is currently working towards the Ph.D. degree with Dept. of Electrical and Electronic Engineering, The University of Hong Kong (HKU), Hong Kong. His recent research interests include edge intelligence and distributed sensing in 6G wireless networks. He was a recipient of Hong Kong Ph.D. Fellowship.

**Xiaojuan Qi** (xjqi@eee.hku.hk) is an assistant professor in the Department of Electrical and Electronic Engineering, the University of Hong Kong. Her research interests include computer vision, artificial intelligence, and deep learning. She was an area chair for CVPR 2023, WACV 2023, AAAI 2021, CVPR 2021, and ICCV 2021 and will be an area chair for NeurIPS 2023.